\documentclass[%
reprint,
 amsmath,amssymb,
 aps,
pra,
]{revtex4-2}

\usepackage{graphicx,xcolor}
\graphicspath{ {./maintext_images/} }
\usepackage{dcolumn}
\usepackage{bm}
\usepackage{braket}
\usepackage[version=4]{mhchem} 

\usepackage{float}      
\usepackage{placeins}   

\newcommand{\HH}{\ce{H2}}
\newcommand{\COO}{\ce{CO2}}
\newcommand{\NNO}{\ce{N2O}}
\newcommand{\NOO}{\ce{NO2}}
\newcommand{\cm}{cm$^{-1}$}
\newcommand{\pHH}{\textit{para}-\ce{H2}}
\newcommand{\oHH}{\textit{ortho}-\ce{H2}}
\newcommand{\ortho}{\textit{ortho}}
\newcommand{\para}{\textit{para}}

\begin{document}

\preprint{}

\title{Environment-imposed selection rules for nuclear-spin conversion of H$_2$ in molecular crystals}

\author{Nathan McLane}
\email{nmclane@umd.edu}
\affiliation{%
 Institute for Physical Science and Technology, \\ University of Maryland, College Park, MD 20742, USA
}%

\author{LeAnh Duckett}
\affiliation{%
 Department of Chemistry and Biochemistry, \\ University of Maryland, College Park, MD 20742, USA
}%

\author{Leah G. Dodson}
\email{ldodson@umd.edu}
\affiliation{%
 Department of Chemistry and Biochemistry, \\ University of Maryland, College Park, MD 20742, USA
}%

\date{\today}

\begin{abstract}
Nuclear-spin conversion in molecular hydrogen is governed by strict symmetry rules that typically require magnetic fields or catalytic surfaces to break. Here we demonstrate that the intrinsic tensor composition of a non-magnetic molecular crystal field can impose and relax these rules without external fields. High-resolution infrared spectra of \HH{} in crystalline \COO{} reveal large rank-2 (quadrupolar) crystal-field splittings of the $m$ sublevels, while nuclear-spin conversion occurs only through $\Delta m = 0$ channels. Replacing \COO{} with polar \NNO{} introduces rank-1 (dipole) components that partially open $\Delta m \neq 0$ pathways, while incorporation of paramagnetic \NOO{} fully lifts the restriction. These results establish a direct correspondence between crystal-field tensor rank and nuclear-spin dynamics, introducing a general symmetry-based framework for designing and controlling spin-isomer populations and quantum-state connectivity in molecular solids.
\end{abstract}

\maketitle


The preparation and control of quantum states are central challenges in modern physics with implications ranging from quantum computing to energy storage \cite{gaita-arino_molecular_2019,wasielewski_exploiting_2020,zhou_quantum_2024}. The governing feature of quantum-state selection is the system symmetry, making symmetry one of the most powerful conceptual tools in molecular and condensed-matter physics \cite{ivanov_quantum-induced_2010,yu_piezoelectricity_2020,zhu_symmetry-driven_2025}. The reduction of complex phenomena in many-body physics to group theoretical elements of the system allows a simplified interpretation of the system, and using methods such as the Wigner-Eckart theorem allows analysis of the selection rules and corresponding quantum dynamics of interacting matter \cite{wigner_gruppentheorie_1931}. These selection rules can apply to specific spectroscopic transitions, govern how the states of a system evolve, or define which channels are closed. The simplest neutral molecular example of state dynamic restriction based on symmetry properties is molecular hydrogen (\HH{}). \HH{} is comprised of two nuclei containing fermionic protons that allow the molecule to be populated into two distinctive nuclear-spin states, \oHH{} and  \pHH{}. Given the symmetry requirements of the total wavefunction, there is no coupling of \ortho{} and \para{} states, consequently often resulting in non-equilibrium distributions of a given state \cite{herzberg_infrared_1946,herzberg_molecular_1950}. To break the symmetry inherent in this system, anisotropic mechanisms have to be built in to induce conversion: paramagnet doping, surface interactions, and matrix confinement all suitably relax symmetry restrictions \cite{nielsen_paramagnetic_1967,ilisca_electron_2014,fedders_nuclear-spin-lattice_1979,yamakawa_fast_2020,senadheera_hydrogen_2008}. When \HH{} is confined in a crystalline environment, the coupling between its spin states depends critically on the symmetries of both the guest (\HH{}) and the host lattice, leading to a range of exotic dynamical behaviors. 

Previous studies have revealed rich symmetry-driven phenomena in confined hydrogen systems. For example, \HH{} encapsulated in \ce{C60} exhibits strong translation–rotation coupling, while \HH{} embedded in the C$_2$ phase of ice shows a pronounced splitting of the $j = 1$ \ortho{} state \cite{mamone_theory_2011,mamone_rotor_2009,yildirim_rotational_2002,di_cataldo_giant_2024}. These works primarily focused on the static properties of confined \HH{}; however, direct observation of time-dependent nuclear-spin conversion dynamics has remained a challenge. In particular, experiments with \ce{C60} could not access these dynamics due to weak overlap between the nuclear-spin states, and similar temporal studies have yet to be carried out for the C$_2$ hydrogen–ice systems that exhibited such large splittings.

\begin{figure*}
\includegraphics[width=\linewidth]{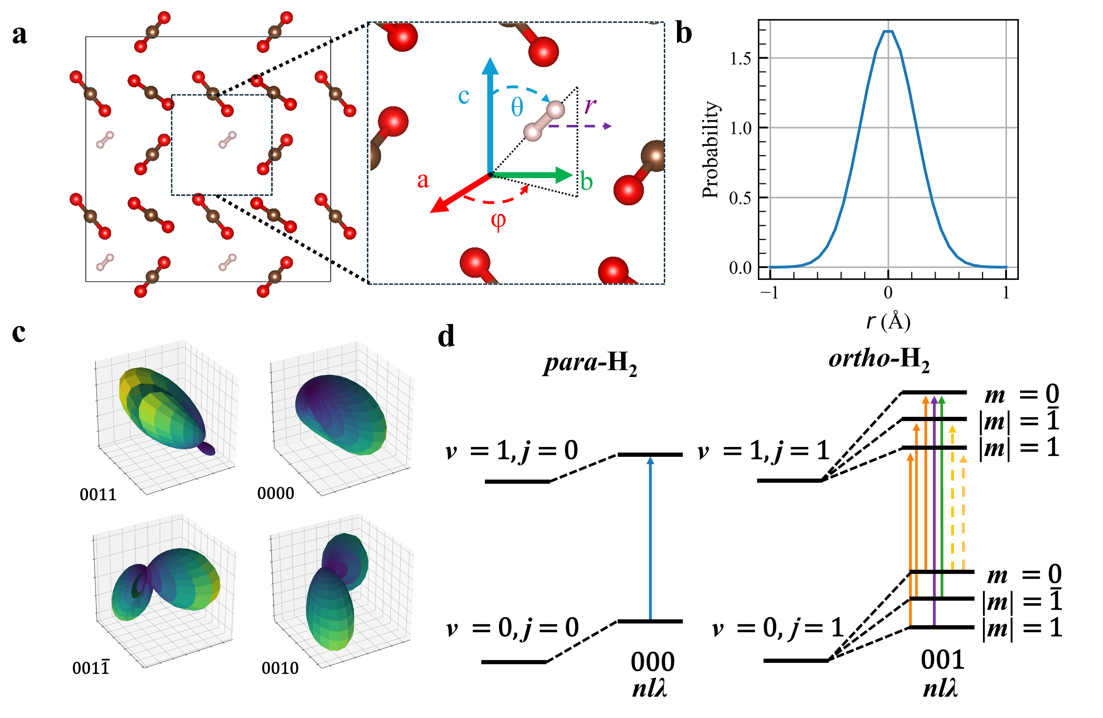}%
\caption{\label{fig:overview} \textbf{The \COO{} crystalline lattice and \HH{} axis definitions with wavefunctions and energies of the perturbed eigenstates.} \textbf{a,} \HH{} trapped in crystalline \COO{} experiences coupling between translational (as a function of $r$) and rotational (as a function of $\theta,\phi$) energies. \textbf{b,} The four lowest-energy translational eigenfunctions have identical probability densities stemming from the harmonic potential. \textbf{c,} Reduced probability density plots of the angular wavefunctions of the four lowest-energy states (shown with state designations $nl\lambda m)$ resemble the spherical harmonics. \textbf{d,} The crystal field of the lattice perturbs the eigenenergies with respect to the gas phase in both \pHH{} and \oHH{} and lowers the degeneracy for $\lambda = 1$ states. Solid arrows indicate rovibrational spectroscopic transitions that are allowed in the perturbed state. Dashed arrows are allowed but suppressed. Four spectroscopic lines are therefore expected in the rotational-vibrational spectrum. }
\end{figure*}

In this work, we demonstrate an environment-imposed selection rule for nuclear-spin conversion in non-magnetic hosts through experiments and theoretical verification of \HH{} confined in crystalline lattices, Fig.~\ref{fig:overview}a. The crystal field produced by diamagnetic \COO{} contains primarily rank-0 and rank-2 tensor components that result in large splitting of the \oHH{} magnetic sublevels; these states are spectroscopically distinguishable in infrared absorption measurements. The lattice has limited rank-1 content, prohibiting nuclear-spin conversion for $|\Delta m| = 1$, while coupling \ortho{} and \para{} states with $\Delta m = 0$ through the field's rank 0/2 elements. These observations are verified using control studies of crystalline \NNO{}, and by adding a paramagnetic impurity \NOO{}.

\section{Results}
The rotational-vibrational structure of \HH{} is well-described based on its fundamental vibrational frequency at 4161 \cm{} and rotational eigenenergies dictated by the rigid rotator with rotational constant $2B \approx 120$ \cm{} \cite{fitzgerald_quantum_2010,fitzgerald_quantum_2008,stoicheff_high_1957}. In confinement, the Hamiltonian acquires additional potential energy terms that depend on the orientation ($\theta,\phi$) of the hydrogen molecule within its local environment and its translational motion (as a function of $r$) \cite{di_cataldo_giant_2024,pauling_rotational_1930,yildirim_quantum_2003,yildirim_rotational_2002,mamone_rotor_2009,mamone_theory_2011,cross_does_2001,matanovic_hydrogen_2012}. These perturbations can be expressed as a sum of spherical-tensor operators $T^{(k)}_q$ that represent the rank-$k$ components of the surrounding crystal field \cite{nielsen_paramagnetic_1967}. Coupling between translational and rotational states results in a total angular momentum quantum number $\Lambda=L+J$ and corresponding eigenstates labeled by $\lambda$ \cite{mamone_theory_2011,mamone_rotor_2009,ye_h2_2013}. The anisotropic components of the crystal field lift the degeneracy of $\lambda$, splitting the energies of the magnetic sublevels $m$ (the projections of $\Lambda$ along the quantization axis) for \oHH{} \cite{di_cataldo_giant_2024}. 

We assumed the effect of the crystalline lattice perturbs the rigid-rotor Hamiltonian through the addition of a potential term
\begin{equation}
V(r,\theta,\phi) = V(r)
+V(\theta,\phi) 
+  V(r)V(\theta,\phi),
\end{equation}
where $V(r)$ is the isotropic translational confinement of \HH{} and $V(\theta,\phi)$ is the angular potential of the local crystal field acting on the rotor. The coupling term $V(r)V(\theta,\phi)$ mixes translational and rotational motion, introducing cross terms between different tensor ranks and enabling small splittings \cite{mamone_theory_2011,mamone_rotor_2009,yildirim_quantum_2003,yildirim_rotational_2002, cross_does_2001}. 
\begin{figure}[h]
\includegraphics[width=86mm]{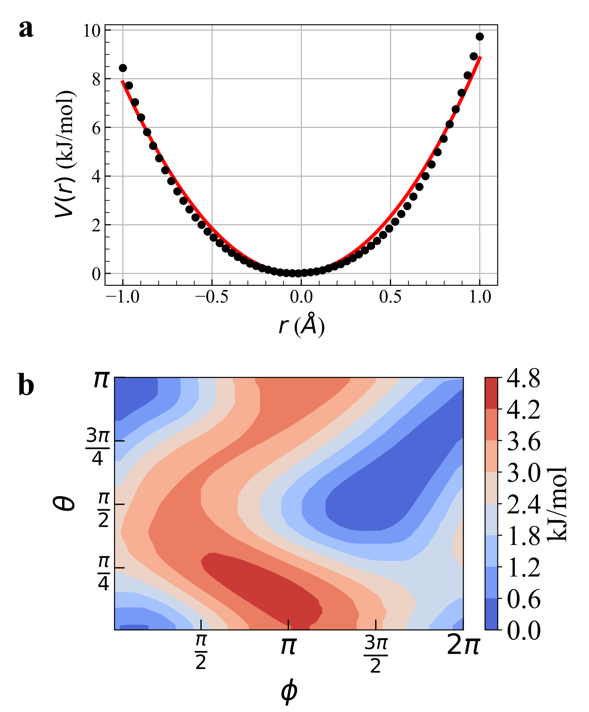}%
\caption{\label{fig:PES-CO2} \textbf{Potential energy plots of \HH{} confined in solid \COO{}.} \textbf{a,} Radial potential $V(r)$ for \HH{} confined in \COO{} as a function of the translational distance $r$ from equilibrium. The black dots are the calculated points and the red line is a fit assuming a harmonic potential. \textbf{b,} Angular potential $V(\theta,\phi)$ for \HH{} in \COO{} with respect to the two rotational degrees of freedom, $\theta$ and $\phi$. The angular potential clearly demonstrates the azimuthal anisotropy that results in splitting of all three $m$ sublevels, and the large field gradients contribute significant rank-2 contributions that enable $\Delta m= 0$ \ortho{}-to-\para{} nuclear-spin conversion.}
\end{figure}

For simplicity we considered a primitive \COO{} unit cell, based on crystallographic information for the pure system \cite{grazulis_crystallography_2012}, with a single \HH{} molecule occupying a substitutional site. The potential functions $V(r)$ and $V(\theta,\phi)$ were computed using the CP2K package on the relaxed geometry obtained for the ground state at the PBE0 level (including an empirical dispersion correction) and TVZ2P basis with a plane wave cutoff of 600 Ry \cite{kuhne_cp2k_2020,adamo_toward_1999}. The translational potential $V(r)$ in Fig.~\ref{fig:PES-CO2}a is harmonic, whereas the angular potential $V(\theta,\phi)$ in Fig.~\ref{fig:PES-CO2}b exhibits significant anisotropy in the two-dimensional surface. Analytical fits to these potentials were then used to numerically obtain energies and wavefunctions (Fig.~\ref{fig:overview}b-c of the perturbed states. Further details may be found in the Methods.

Eigenenergies for $j=$ even (\pHH) and $j=$ odd (\oHH) are diagrammed in Fig.~\ref{fig:overview}d, showing the splitting of the $\ket{\lambda m}$ states for $\lambda=1$. Energies of all perturbed states reported in Table~\ref{tab:rotE} are destabilized with respect to the free-rotor states. Higher-energy states are also reported in Supplementary Table S1. The $m$ quantum numbers for \oHH{} are identified by inspection of the reduced probability density plots with respect to the $z$ axis aligned to the $C_3$ symmetry axis of the \COO{} local environment. Because the local crystal field breaks azimuthal symmetry, the two $|m|=1$ states are non-degenerate and are labeled arbitrarily $m=1$ and $\bar{1}$. While there remains only a single rovibrational transition in the \pHH{} manifold, $\Delta m = 0$ and $|\Delta m| = 1$ transitions are possible during vibrational excitation of \oHH{} in the Q-branch manifold \cite{fitzgerald_quantum_2008, kong_first-principles_2011,fitzgerald_quantum_2010,van_kranendonk_crystal_1962}. Solid lines in Fig.~\ref{fig:overview}d indicate all expected transitions, whereas dashed lines indicate transitions that are symmetry suppressed \cite{kong_first-principles_2011}. Four observable spectroscopic transitions are therefore predicted: one transition for molecules in the \para{} nuclear-spin state; and three non-degenerate transitions for molecules with $j=1$: one three-fold degenerate transition for $\Delta m = 0$ (orange arrows) and two transitions corresponding to $|\Delta m| = 1$ (green and purple arrows).
\begin{figure}[h]
\includegraphics[width=86mm]{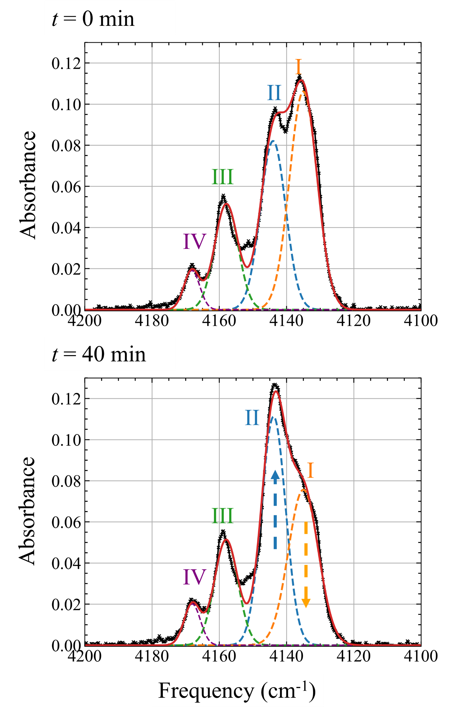}%
\caption{\label{fig:H2-CO2} \textbf{The Q$_1$ spectral region of 1\% \HH{} confined in \COO{} at 10 K.} \textbf{top,} $t = 0$, and \textbf{bottom,} $t = 40$ min. Experimental data are shown in black. The dashed lines correspond to the individual fitted Gaussian peaks, and the red line is the linear combination of all fitted peaks. Peak I (orange dashed line) is the absorption peak of \oHH{} molecules from any $m$ sublevel with $\Delta m = 0$, and is seen to decrease with time as molecules originally in $\ket{10}$ convert to $\ket{00}$. Peak II (blue dashed line) corresponds to absorption by \pHH{} molecules, with a corresponding increase in intensity with time as \ortho-to-\para{} conversion occurs. Peaks III and IV (green and purple dashed lines) are assigned to absorption by \oHH{} initially in an $|m|=1$ state ($m=\pm1$ split, labeled $1$ and $\bar{1}$ by energy) that are weakly excited by a $|\Delta m|=1$ transition to the $\ket{10}$ state in the vibrational excited state. They do not evolve with time due to the host-conditioned selection rule for nuclear-spin conversion.}
\end{figure}

\begin{table}[t]
\caption{\label{tab:rotE}%
Calculated energies (in \cm{}) of the perturbed system.
}
\begin{ruledtabular}
\begin{tabular}{ccccc}
{} & 
\multicolumn{2}{c}{\COO{}} & 
\multicolumn{2}{c}{\NNO{}} \\
State $\ket{\lambda m}$ & 
Energy & 
$\Delta E$\footnotemark[1]  & 
Energy & 
$\Delta E$\footnotemark[1] \\
\colrule
        $\ket{11}$ & 369.2 & 0 &  284.3 &  0        \\
        $\ket{00}$ & 386.0 & 16.8 & 312.9 & 28.6  \\
        $\ket{1\bar{1}}$ & 406.1 & 36.9 & 288.2 & 3.9 \\
        $\ket{10}$ & 430.1 & 60.9 & 322.2  & 38.6 \\
\end{tabular}
\end{ruledtabular}
\footnotetext[1]{Relative to $\ket{10}$ state.}
\end{table}
\begin{table}[t]
\caption{\label{tab:peaks}%
Summary of observed rovibrational transitions of \HH{} in a \COO{} or \NNO{} ice (\cm{}) identified by $\ket{\lambda m}\rightarrow\ket{\lambda' m'}$ during excitation from vibrational state $v=0$ to $v'=1$. 
}
\begin{ruledtabular}
\begin{tabular}{cccc}
Peak &
Transition  & 
\HH{} in \COO{}  & 
\HH{} in \NNO{} \\
\colrule
    I & $\ket{1m}\rightarrow\ket{1m}$\footnotemark[1]  & 4134.4    &  4132.6\footnotemark[3] \\
    II & $\ket{00}\rightarrow\ket{00}$   & 4143.9    &  4143.6  \\
    III & $\ket{1\bar{1}}\rightarrow\ket{10}$\footnotemark[2]   &  4158.1   & 4132.6\footnotemark[3]   \\
    IV & $\ket{11}\rightarrow\ket{10}$\footnotemark[2]  &  4168.2   & 4150.9   \\
\end{tabular}
\end{ruledtabular}
\footnotetext[1]{For $m=0,1,\bar{1}$, with $\Delta m=0$.}
\footnotetext[2]{$|\Delta m|=1$.}
\footnotetext[3]{Assumed to be unresolved.}
\end{table}

The rovibrational experimental spectrum of normal hydrogen trapped in frozen carbon dioxide (1\% \HH{} in \COO{} at 10 K) confirms the prediction of four optical transitions, with four distinct peaks appearing in the Q$_1$ spectrum, shown in Fig.~\ref{fig:H2-CO2}. Peak centers, obtained from Gaussian fits to the experimental spectrum, are listed in Table~\ref{tab:peaks}. The transition energies match those predicted by calculations. The lowest-energy peak (peak I, orange dashed line) corresponds to \oHH{} transitions from all three initial $m$ states while obeying the strict $\Delta m = 0$ selection rule. The second-lowest-energy peak (peak II, blue dashed line) is the only peak corresponding to excitation of \pHH{} molecules, appearing 9.5 \cm{} higher in energy than the $\Delta m = 0$ \ortho{} peak (predicted 17 \cm{}). The final two peaks (peaks III and IV, green and purple dashed lines) correspond to transitions of \oHH{} molecules starting in the $|m|=1$ state prior to vibrational excitation and converting to $m'=0$ in the vibrational excited state ($|\Delta m| = 1$). These two peaks are observed 23.7 and 33.8 \cm{} above the $\Delta m = 0$ peaks---qualitatively in line with prediction---and are weak as expected \cite{kong_first-principles_2011}. 

The time evolution of the experimental spectrum provides additional verification for assignments of these transitions: the intensity of peak I decreases from $t=0$ to $40$ minutes, whereas the intensity of peak II increases with time, implying a population transfer between the corresponding states. We expect initially deposited \oHH{} (room temperature \ortho-to-\para{} ratio = 3:1) to be driven by thermodynamics to convert to \pHH{} through weak tensor couplings that mediate nuclear-spin conversion. In the non-magnetic host, only the lowest-rank ($k=0,2$) components of the crystal-field tensor are appreciable, restricting the allowed transitions to $\Delta m=0$. Therefore, the loss of intensity in peak I confirms that it originates from \ortho{} molecules whose $m=0$ sublevel can couple to \pHH{} via these even-rank tensor terms, and conversely the increase in peak II corroborates its assignment as belonging to the \para{} manifold. 

The time invariance of peaks III and IV, along with the persistence of peak I---whose intensity plateaus after 40 minutes---demonstrates the absence of rank-1 tensor content in the host and confirms the resulting host-conditioned nuclear-spin-conversion selection rule. These three transitions are assigned to rovibrational excitation of molecules in the \ortho{} nuclear-spin state. At 10 K, the expected equilibrium \ortho-to-\para{} ratio is approximately $10^{-7}$, thus, if nuclear-spin conversion channels were available, all \HH{} molecules would relax to the \para{} state. While nuclear-spin conversion is formally forbidden in the gas phase, this process can occur in non-magnetic solids through even-rank rotational-spin, spin-lattice, and spin-spin dipolar interactions \cite{pileio_relaxation_2010,fedders_nuclear-spin-lattice_1979}. \oHH{} molecules initially in the $\ket{10}$ state can convert to \pHH{} ($\ket{00}$) through $\Delta j = -1$, $\Delta m = 0$ transitions driven by rank-0 and rank-2 perturbations of the crystal field. The observed population transfer between peak I (\oHH{}) and II (\pHH{}) corroborates this framework. The higher-energy peaks III and IV, assigned to molecules initially in the $\ket{11}$ and $\ket{1\bar{1}}$ states, require both $\Delta j = -1$ and $|\Delta m| = 1$; their invariance with time confirms that these channels remain closed in the absence of rank-1 contributions. These substates also contribute weakly to the lowest-energy rovibrational transition (peak I), and the non-zero asymptotic behavior of this peak further corroborates the closure of these channels. 

\section{Discussion}

\begin{figure}[h]
\includegraphics[width=86mm]{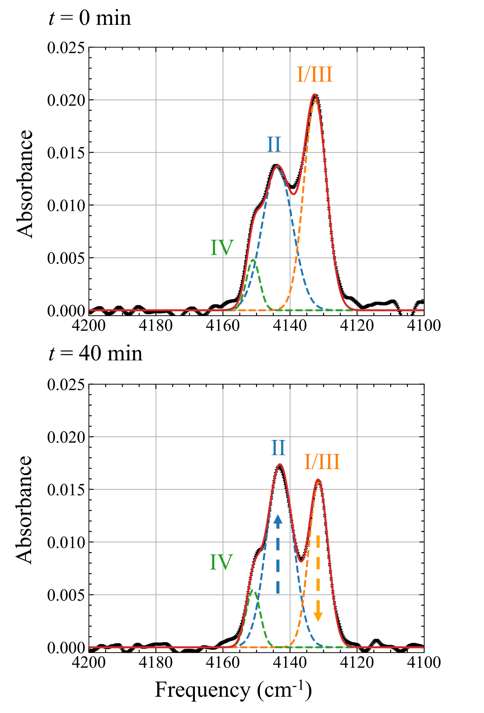}%
\caption{\label{fig:H2-N2O} \textbf{The Q$_1$ spectral region of 1\% \HH{} in \NNO{} at 10 K.} \textbf{top,} $t = 0$, and \textbf{bottom,} $t = 40$ min. Experimental data are shown in black. The dashed lines correspond to the individual fitted Gaussian peaks, and the red line is the linear combination of all fitted peaks. Peak I/III (orange dashed line) has contributions from nearly-degenerate transitions following the $\Delta m = 0$ spectroscopic selection rule and the weak $\ket{1\bar{1}} \rightarrow \ket{10}$ excitation. This peak decreases with time as molecules originally in $\ket{10}$ convert to $\ket{00}$ by the host-conditioned nuclear-spin conversion selection rule. Peak II (blue dashed line) again corresponds to absorption by \pHH{} molecules, with a corresponding increase in intensity with time as \ortho-to-\para{} conversion occurs. Peak IV (green dashed line) originates from \oHH{} weak $\ket{11}\rightarrow\ket{10}$ spectroscopic transitions. It does not evolve with time due to the host-conditioned selection rule for nuclear-spin conversion.}
\end{figure}

\begin{figure}[h]
\includegraphics[width=86mm]{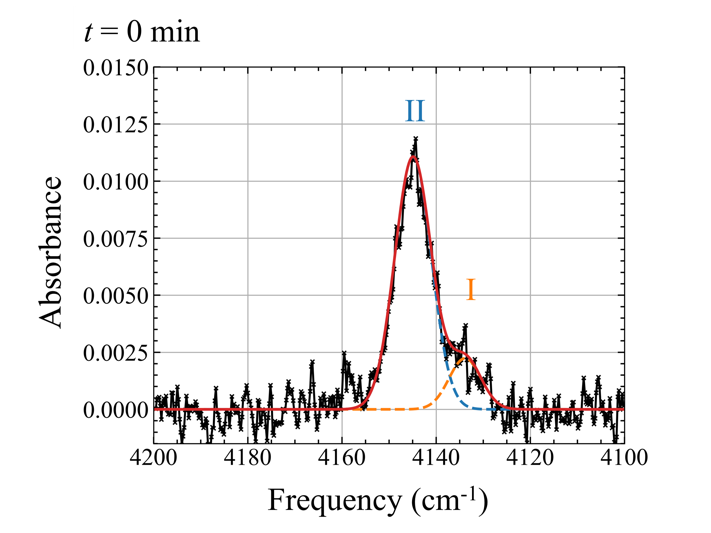}%
\caption{\label{fig:H2-NO2-CO2} \textbf{The Q$_1$ branch of 1\% H$_2$ co-trapped with 3\% \NOO{} in \COO{} ice at 10 K.} Experimental data ($t=0$ min) are shown in black. The dashed lines correspond to the individual fitted Gaussian peaks, and the red line is the linear combination of all fitted peaks. Peak I (orange dashed line) is residual \oHH{} molecules. Peak II (blue dashed line) corresponds to absorption by \pHH{} molecules, showing that nearly all of the \oHH{} molecules have converted to \pHH{} in the presence of a paramagnetic impurity.}
\end{figure}

Comparison with other host systems highlights how the confined environment dictates the perturbative nature and defines spectroscopic selection rules, the magnitude of the $m$ energy splitting, and the channels available for nuclear-spin conversion. Amorphous \COO{} films lack the crystalline field needed to produce rank-2 tensor elements that clearly split $m$-substates, while fast nuclear-spin conversion may occur through $|\Delta m| = 1$ mixing \cite{yamakawa_fast_2020}. In endohedral \ce{C60} the local crystalline lattice, although sharing the same space group as \COO{}, is nearly isotropic with very small distortions that contribute rank-2 content to create small splittings in the sublevels. Based on measured lifetimes, \HH{} entrapped in \ce{C60} is assumed to have no open channels to nuclear-spin conversion \cite{xu_coupled_2009,ye_h2_2013,mamone_theory_2011,mamone_rotor_2009,yildirim_rotational_2002,kohama_rotational_2009}. In metal-organic frameworks \cite{fitzgerald_orientational_2015,kong_first-principles_2011,fitzgerald_quantum_2008,matanovic_hydrogen_2012}, the metal sites will contribute rank-2 content, while absent rank-1 contributions, similarly leading to relatively large splittings in sublevel energies but without the large crystal-field gradients present in crystalline \COO{} that enable nuclear-spin conversion. Crystalline water ice (C$_2$ phase) is a host with large rank 0/1/2 contributions, leading to large energy splittings and fast nuclear-spin conversion \cite{di_cataldo_giant_2024,senadheera_hydrogen_2008}. In the present work, complementary experiments replace \COO{} with its isoelectronic analog \NNO, which also forms the \ce{Pa$\bar{3}$} space group in low-temperature crystals \cite{solodovnik_structure_2009}. \NNO{} will have reduced rank-2 content compared with \COO{} due to its smaller quadrupole, and the weak dipole moment adds an element of rank-1 coupling. We obtained $V(r)$ and $V(\theta,\phi)$ for crystalline \NNO{} (see Supplementary Information section S2 for further details) and used them to obtain the eigenfunctions and eigenenergies for this confined \HH{} system and compared the expected optical transitions with spectra obtained for 1\% \HH{} in \NNO{} at 10 K (Fig.~\ref{fig:H2-N2O}). The $m$-state energy splittings are still large (Table~\ref{tab:rotE}), but not as large as those of \COO{}; verifying the importance of the dominant rank-2 tensor elements in \COO. The degeneracy of the \NNO{} system is only partially lifted, with the two lowest-energy substates of \oHH{} remaining nearly degenerate. Transitions from these substates are observed as one single peak (peak \NNO-I/III) listed in Table~\ref{tab:peaks}.
The presence of a paramagnetic species like \NOO{} introduces rank-1 tensor elements---at only 3\% \NOO{} impurity in a 1\% \HH{} in \COO{} ice at 10 K, all of the peaks assigned to \oHH{} disappear as the $|\Delta m|=1$ channels open, and we are left with the \pHH{} peak; Peak II in Fig.~\ref{fig:H2-NO2-CO2}. The \NNO{} and \NOO{} control experiments combine with the results of the present work to create a complete picture of the impact of the crystal field in molecular crystal lattices on the quantum dynamics of \HH{}.

\begin{figure}[h]
\includegraphics[width=86mm]{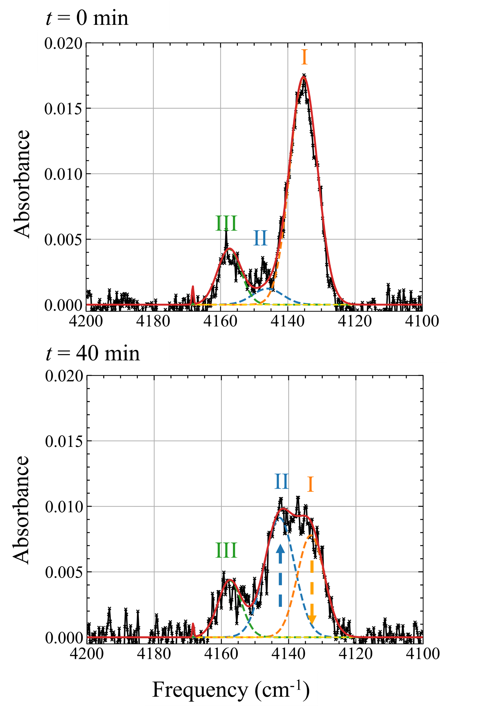}%
\caption{\label{fig:highH2-CO2} \textbf{The Q$_1$ branch of \HH{} confined at higher concentrations (10\%) \HH{} in \COO{} ice at 10 K.} \textbf{top,} $t = 0$, and \textbf{bottom,} $t = 110$ min. Experimental data are shown in black. The dashed lines correspond to the individual fitted Gaussian peaks, and the red line is the linear combination of all fitted peaks. Peak I (orange dashed line) is the absorption peak of \oHH{} molecules from any $m$ sublevel with $\Delta m = 0$, and is seen to decrease with time as molecules originally in $\ket{10}$ convert to $\ket{00}$. Its intensity ratio to the \pHH{} peak (peak II, blue dashed line) is much larger at higher \HH{} concentration due to preferential binding of \ortho{} molecules. Compared with Fig.~\ref{fig:H2-CO2}, there is a diminishment in peaks III and IV, which is opposite of the expected behavior if these peaks were due to site effects.}
\end{figure}

While site effects are uncommon for low-concentration matrix-isolated diatomic molecules \cite{andrews_chemistry_1989}, the possibility that peaks III and IV may be caused by \HH{} molecules trapped in geometrically distinct sites (e.g. interstitial sites of the crystalline \COO) is tested using concentration-dependent studies. Increasing the guest-to-host ratio should increase the probability of guest molecules occupying energetically disfavored sites, leading to a corresponding increase in the observed absorption intensity of these sites. However, experiments conducted with 10\% \HH{} in \COO{} reveal the opposite trend, with Peak IV disappearing entirely (Fig.~\ref{fig:highH2-CO2}). This is evidence to the contrary: that the \HH-\HH{} interactions begin to increase at higher concentrations, causing notably, an induced dipole canceling mechanism termed the ``cancellation effect'' \cite{gush_infrared_1960,van_kranendonk_crystal_1962}, but no increase or introduction of new absorption features. These spectra also imply preferential binding of \oHH{} during the preliminary deposition process, which is revealed through the initial absence peak II (\pHH{}) immediately upon deposition. This preferential binding has been observed in other ices and van der Waals complexes \cite{grinev_separation_2007,buch_preferential_1993}. The peak II intensity is ultimately recovered as \oHH{} $\ket{10}$ species in peak I convert to the \para{} form. 

\section{Conclusions}
In summary, crystalline \COO{} acts as a quantum-state filter that enforces a host-conditioned selection rule for nuclear-spin conversion of \HH{}. The nonmagnetic lattice provides strong rank-2 (quadrupolar) anisotropy but negligible rank-1 (vector) content, yielding large splittings of the $m$ sublevels yet restricting nuclear-spin conversion to the $\Delta m = 0$ channel. Replacing \COO{} with polar \NNO{} introduces weak rank-1 components that partially relax this restriction, while paramagnetic \NOO{} fully opens all spin-conversion channels. Together these results establish a direct correspondence between the tensor composition of the local field and the available nuclear-spin pathways. Beyond hydrogen, this ``tensor-engineered'' control of state connectivity offers a design principle for manipulating quantum-state populations in simple molecular solids.
Such direct and coherent interconversion between pure spin states opens new possibilities for realizing molecular qubits, probing local field anisotropies, and symmetry-based control of quantum-state evolution in condensed environments.

\begin{acknowledgments}
This material is based upon work supported by the U.S. Department of Energy, Office of Science Early Career Research Program, Office of Basic Energy Sciences under Award Number DE-SC0024262. The authors acknowledge the University of Maryland supercomputing resources (http://hpcc.umd.edu) made available for conducting the research reported in this paper.
\end{acknowledgments}

\appendix

\section{Methods}
\subsection{\label{appX:exp}Experimental.}
The matrix-isolation spectroscopy instrument has been described in previous work \cite{hockey_weakly_2022,hockey_matrix-formation_2024}. In short, a mixture of gases is deposited on a 1-in diameter ZnSe substrate mounted at the base of a CTI Model 22 cryocooler. A Lakeshore Model 331 temperature controller monitors and controls the temperature through both a silicon diode and resistive heaters. The cryocooler extends through a differentially pumped rotary platform that enables the attached substrate to rotate between the deposition axis and the spectroscopy axis without breaking vacuum. The substrate either faces the dosing valve with gas colliding with the cold surface at a slight angle or it faces the direction of the spectrometer so that the IR light transmits directly through the substrate and any accumulated sample. 

The IR spectra are collected using a Nicolet iS50 Fourier Transform Infrared (FTIR) Spectrometer after the light makes a single pass through the matrix and ZnSe optic. After passing through, the IR light arrives at an external MCT-A detector. In its current configuration the spectrometer is equipped with a KBr beamsplitter; combined with the MCT-A detector, the available spectral range is 7400 to 600 \cm{}. The spectra are recorded with a resolution of 0.5 \cm{} and a data spacing of 0.241 \cm{}. Spectroscopic fitting is done using an in-house Python script fitting each specific ro-vibrational transition to a Gaussian function.

In the present work, mixed matrices are deposited directly onto the cold substrate by flowing gas mixtures directly from prepared cylinders into the chamber. Various concentrations of molecular hydrogen (\HH{}) ranging from 1\% to 10\% are prepared by diluting pure \HH{} (Airgas, UHP) with excess \NNO{} (Airgas, USP) or \COO{} (Airgas, Industrial Grade) in a stainless-steel cylinder to total pressures of 1500 Torr, using manometric techniques. The gaseous mixtures are then attached to a gas manifold directly connected to the matrix-isolation chamber allowing deposition on the cold ZnSe substrate held at 10 K. Deposition rates are controlled through a VAT dosing valve, ranging from 0.225 mtorr L s$^{-1}$ to 7,500 mtorr L s$^{-1}$ for a total deposition time of 30 min. Crystallinity of the \COO{} matrix is verified by the presence of Davydov splitting in the infrared spectrum, shown in Supplementary Fig. S1 \cite{escribano_crystallization_2013,isokoski_highly_2013}. 

\subsection{\label{appX:theory}Theoretical.}

The energy eigenvalues, $E$, of the freely rotating quantum rigid rotor \HH{} are obtained by solving the Schr{\"o}dinger equation in spherical coordinates

\begin{equation}
\frac{-\nabla^2 \psi(\theta,\phi)}{2 \mu R^2}  =
E \psi(\theta,\phi) \label{appEqnA}
\end{equation}

\noindent where $\mu$ is the reduced mass of molecular hydrogen and $R$ is the radial distance between the two hydrogen atoms.

In the case that \HH{} is trapped in a confined environment such as a molecular ice, the Hamiltonian acquires additional potential energy terms that are dependent on the hydrogen molecule's orientation in its local environment ($\theta , \phi$) and trapping potential with respect to its translational motion ($r$) in the confined space \cite{pauling_rotational_1930,yildirim_rotational_2002,mamone_theory_2011,mamone_rotor_2009,cross_does_2001}. Other couplings can occur with the rotational and vibrational states, but to a good approximation the rotational-vibrational couplings of \HH{} in confined environments are small due to the molecule's large rotational constant \cite{yildirim_rotational_2002,pauling_rotational_1930}. This small effect enables calculation of the translational-rotational energies and corresponding states of \HH{} in a confining environment at its equilibrium geometry with additional potentials included, producing the total 5D Hamiltonian \cite{mamone_rotor_2009, sebastianelli_hydrogen_2007,mamone_theory_2011}

\begin{equation}
H = -\frac{\hbar^{2}}{2M} 
\left( \frac{\partial^{2}}{\partial x^{2}} 
     + \frac{\partial^{2}}{\partial y^{2}} 
     + \frac{\partial^{2}}{\partial z^{2}} \right) 
+ B \hat{j}^{2} 
+ V(x,y,z,\theta,\phi). \label{appEqnB}
\end{equation}

\noindent In this Hamiltonian $M$ is the total mass of \HH{}, $B$ is the rotational constant of \HH{} and $\hat{j}^{2}$ is the angular momentum operator for \HH{}. $V(x,y,z,\theta,\phi)$ is the 5D potential applied to the \HH{} molecule by the surrounding medium impacting the energies of the translational and rotational energies as well allowing as coupling between the two states. Depending on the symmetry of the entrapping environment,  the translational motion of the molecule can be approximated as isotropic, reducing the overall Hamiltonian to a three-dimensional problem: 

\begin{equation}
H = -\frac{\hbar^{2}}{2M} 
\left( \frac{\partial^{2}}{\partial r^{2}} 
 \right) 
+ B \hat{j}^{2} 
+ V(r,\theta,\phi). \label{appEqnC}
\end{equation}

Assuming an isotropic symmetry for the translational motion simplifies the potential function to the form

\begin{equation}
V(r,\theta,\phi) = V(\theta,\phi) 
+V(r)
+  V(r)V(\theta,\phi), \label{appEqnD}
\end{equation}

\noindent (Refs.~\cite{mamone_rotor_2009,mamone_theory_2011, cross_does_2001}) where the first two terms  $V(r)$ and $V(\theta,\phi)$ represent the translational potential felt by \HH{} moving in one direction $(r)$ and the angular potential of rotating \HH{} in its two rotational degrees of freedom $(\theta,\phi)$, respectively. The third term is a coupling term that couples the translational and rotational states inducing mixing between the two states. This theoretical framework is based on previous work that demonstrated the translational-rotational dynamics of incarcerated \HH{} molecules in \ce{C60} systems \cite{mamone_theory_2011,mamone_rotor_2009} and is extended to include the separate angular potential term, $V(\theta,\phi)$,  given that the molecular crystals investigated in the present work impose appreciable angular potentials on the confined \HH{}, similar to \ce{H2O} \cite{di_cataldo_giant_2024}. Using these methodologies is appropriate for the molecular systems in this work since both \COO{} and \NNO{}, in their crystalline forms, form the same confining space group (cubic \ce{Pa$\bar{3}$}) as \ce{C60} \cite{yildirim_rotational_2002,solodovnik_structure_2009}.

The contributions of $V(r)$ and $V(r)V(\theta,\phi)$ require four more quantum numbers--- $n, l, \lambda, m$--- to completely describe the state of the \HH{} molecule.  The quantum number $n$, corresponds to the quantization of translational excitations, and $l$ is the quantization of the angular momentum of \HH{} for a given $n$. The final two quantum numbers $\lambda$ and $m$ are the total angular momentum defined by $\Lambda = L + J$, and the projection of $\lambda$ on the quantization axis, respectively. Given that $V(\theta,\phi)$ is small, the rotational quantum number $j$ is still a good quantum number for $\psi(r,\theta,\phi$) but breaks the inherit $2j + 1$ degeneracy \cite{pauling_rotational_1930,fitzgerald_quantum_2008,yildirim_rotational_2002,di_cataldo_giant_2024}. Without the angular-dependent potential $V(\theta,\phi)$, $\lambda$ has a $2\lambda + 1$  degeneracy; however, including $V(\theta,\phi)$ allows splitting of the $m$ quantum states to unique energy value states. To clearly assign the translational-rotational states of confined \HH{} the overall nomenclature used is $jnl{\lambda}m$ to define all quantum numbers, which is reduced to $\ket{\lambda m}$ in discussing \ortho{} and \para{} states in the text. 

The 3D potential is calculated using computational quantum-chemical methods, producing $V(r,\theta,\phi)$ 
\cite{cross_does_2001,di_cataldo_giant_2024}. Density functional theory calculations are conducted on a single \HH{} molecule substituted into the primitive unit cell of either \COO{} or \NNO{} using the CP2K package to compute $V(r)$ and $V(\theta,\phi)$ \cite{kuhne_cp2k_2020}. Crystallographic information for the pure \COO{} and \NNO{} unit cells from the Crystallography Open Database \cite{grazulis_crystallography_2012} provides a starting point for the cell lattice. A \COO{} (or \NNO{}) molecule from the crystal lattice is replaced with an \HH{} molecule, and a geometry optimization is carried out at the PBE0 level of theory with a TVZ2P basis set, a plane wave cutoff of 600 Ry and relative cutoff of 100 Ry \cite{adamo_toward_1999}. To account for dispersion interactions within the crystal cell, an additional empirical D3 correction is applied \cite{grimme_consistent_2010}. The absence of imaginary frequencies verifies that the optimization reaches a minimum configuration. To compute the 1D potential $V(r)$, the center of mass of \HH{} is translated along one direction in the crystal lattice, starting from the optimized structure, with single-point energies computed as a function of $r$. To calculate $V(\theta,\phi)$, the \HH{} molecule at equilibrium is rotated over a 60 by 60 grid encompassing the range of 0 to $\pi$ and 0 to 2$\pi$ for $\theta$ and $\phi$, respectively.

The corresponding energy eigenvalues and translational-rotational wavefunctions, \textit{E} and $\psi(r,\theta,\phi)$, are obtained by numerically solving the Schr{\"o}dinger equation (Eq.~\ref{appEqnC}). The functional form of $V(r)$ is assumed to be harmonic, and the single-point energy values obtained computationally are fit accordingly. The angular potential $V(\theta,\phi)$ is fit to a linear combination of sines and cosines to reduce the total number of points used in the overall Schr{\"o}dinger solver to a $30 \times 30 \times 30$ grid ($r,\theta,\phi$). Translational-rotational energy eigenvalues and wavefunctions are numerically obtained using a custom-written Python code using the Lanczos algorithm in shift and invert mode to solve for the lowest energies and eigen-states. Initial determination of the accuracy of the algorithm is compared to the analytical results of the rigid rotor model.

\bibliographystyle{unsrturl}   

\begin{thebibliography}{10}

\bibitem{gaita-arino_molecular_2019}
A.~Gaita-Ariño, F.~Luis, S.~Hill, and E.~Coronado.
\newblock Molecular spins for quantum computation.
\newblock {\em Nature Chemistry}, 11(4):301--309, April 2019.
\newblock URL: \url{https://www.nature.com/articles/s41557-019-0232-y}, \href {https://doi.org/10.1038/s41557-019-0232-y} {\path{doi:10.1038/s41557-019-0232-y}}.

\bibitem{wasielewski_exploiting_2020}
Michael~R. Wasielewski, Malcolm D.~E. Forbes, Natia~L. Frank, Karol Kowalski, Gregory~D. Scholes, Joel Yuen-Zhou, Marc~A. Baldo, Danna~E. Freedman, Randall~H. Goldsmith, Theodore Goodson, Martin~L. Kirk, James~K. McCusker, Jennifer~P. Ogilvie, David~A. Shultz, Stefan Stoll, and K.~Birgitta Whaley.
\newblock Exploiting chemistry and molecular systems for quantum information science.
\newblock {\em Nature Reviews Chemistry}, 4(9):490--504, July 2020.
\newblock URL: \url{https://www.nature.com/articles/s41570-020-0200-5}, \href {https://doi.org/10.1038/s41570-020-0200-5} {\path{doi:10.1038/s41570-020-0200-5}}.

\bibitem{zhou_quantum_2024}
Yuanyuan Zhou, Wei Yuan, Mengting Li, Zhenyang Xie, Xiaoyun Song, Yang Yang, Jian Wang, Li~Li, Wei Ding, Wen-Feng Lin, and Zidong Wei.
\newblock Quantum confinement-induced anti-electrooxidation of metallic nickel electrocatalysts for hydrogen oxidation.
\newblock {\em Nature Energy}, 9(10):1297--1309, August 2024.
\newblock URL: \url{https://www.nature.com/articles/s41560-024-01604-9}, \href {https://doi.org/10.1038/s41560-024-01604-9} {\path{doi:10.1038/s41560-024-01604-9}}.

\bibitem{ivanov_quantum-induced_2010}
Sergei~D. Ivanov, Oskar Asvany, Alexander Witt, Edouard Hugo, Gerald Mathias, Britta Redlich, Dominik Marx, and Stephan Schlemmer.
\newblock Quantum-induced symmetry breaking explains infrared spectra of {CH}$_{\textrm{5}}$$^{\textrm{+}}$ isotopologues.
\newblock {\em Nature Chemistry}, 2(4):298--302, April 2010.
\newblock URL: \url{https://www.nature.com/articles/nchem.574}, \href {https://doi.org/10.1038/nchem.574} {\path{doi:10.1038/nchem.574}}.

\bibitem{yu_piezoelectricity_2020}
Jiabin Yu and Chao-Xing Liu.
\newblock Piezoelectricity and topological quantum phase transitions in two-dimensional spin-orbit coupled crystals with time-reversal symmetry.
\newblock {\em Nature Communications}, 11(1):2290, May 2020.
\newblock URL: \url{https://www.nature.com/articles/s41467-020-16058-2}, \href {https://doi.org/10.1038/s41467-020-16058-2} {\path{doi:10.1038/s41467-020-16058-2}}.

\bibitem{zhu_symmetry-driven_2025}
Peide Zhu, Zhixin Liu, Xia Lei, Siru He, Deng Wang, Jie Zeng, Lida Wang, Fei Su, Wenbo Peng, Zheng Liang, Yuxin Sun, Zhiwei Lei, Zhitong Li, Hsien-Yi Hsu, Xu~Pan, Xingzhu Wang, Jingbai Li, Yong Zhang, and Baomin Xu.
\newblock Symmetry-driven engineering of long-range-ordered $\pi$-$\pi$ stacking molecules for high-efficiency perovskite photovoltaics.
\newblock {\em Nature Synthesis}, September 2025.
\newblock URL: \url{https://www.nature.com/articles/s44160-025-00896-3}, \href {https://doi.org/10.1038/s44160-025-00896-3} {\path{doi:10.1038/s44160-025-00896-3}}.

\bibitem{wigner_gruppentheorie_1931}
Eugen Wigner.
\newblock {\em Gruppentheorie und ihre {Anwendung} auf die {Quantenmechanik} der {Atomspektren}}.
\newblock Vieweg+Teubner Verlag, Wiesbaden, 1931.
\newblock URL: \url{http://link.springer.com/10.1007/978-3-663-02555-9}, \href {https://doi.org/10.1007/978-3-663-02555-9} {\path{doi:10.1007/978-3-663-02555-9}}.

\bibitem{herzberg_infrared_1946}
Gerhard Herzberg and Bryce~L. Crawford.
\newblock Infrared and {Raman} {Spectra} of {Polyatomic} {Molecules}.
\newblock {\em The Journal of Physical Chemistry}, 50(3):288--288, March 1946.
\newblock URL: \url{https://pubs.acs.org/doi/abs/10.1021/j150447a021}, \href {https://doi.org/10.1021/j150447a021} {\path{doi:10.1021/j150447a021}}.

\bibitem{herzberg_molecular_1950}
Gerhard Herzberg.
\newblock {\em Molecular {Spectra} and {Molecular} {Structure}. {I}. {Spectra} of {Diatomic} {Molecules}}.
\newblock D. Van Nostrand Company, Inc, Princeton, NJ, 1950.

\bibitem{nielsen_paramagnetic_1967}
Svend~E. Nielsen and John~S. Dahler.
\newblock Paramagnetic catalysis of the ortho-parahydrogen conversion.
\newblock {\em The Journal of Chemical Physics}, 46(2):732--744, January 1967.
\newblock URL: \url{https://pubs.aip.org/jcp/article/46/2/732/80420/Paramagnetic-Catalysis-of-the-Ortho-Parahydrogen}, \href {https://doi.org/10.1063/1.1840734} {\path{doi:10.1063/1.1840734}}.

\bibitem{ilisca_electron_2014}
Ernest Ilisca and Filippo Ghiglieno.
\newblock Electron exchanges in nuclear spin conversion of hydrogen physisorbed on diamagnetic insulators.
\newblock {\em The European Physical Journal B}, 87(10):235, October 2014.
\newblock URL: \url{http://link.springer.com/10.1140/epjb/e2014-50282-2}, \href {https://doi.org/10.1140/epjb/e2014-50282-2} {\path{doi:10.1140/epjb/e2014-50282-2}}.

\bibitem{fedders_nuclear-spin-lattice_1979}
Peter~A. Fedders.
\newblock Nuclear-spin-lattice relaxation times for {H}$_{\textrm{2}}$ in solid nonmagnetic hosts.
\newblock {\em Physical Review B}, 20(7):2588--2593, October 1979.
\newblock URL: \url{https://link.aps.org/doi/10.1103/PhysRevB.20.2588}, \href {https://doi.org/10.1103/PhysRevB.20.2588} {\path{doi:10.1103/PhysRevB.20.2588}}.

\bibitem{yamakawa_fast_2020}
Koichiro Yamakawa, Atsuki Ishibashi, Toshinobu Namiyoshi, Yuichi Azuma, and Ichiro Arakawa.
\newblock Fast nuclear-spin conversion of {H}$_{\textrm{2}}$ trapped and polarized in a {CO}$_{\textrm{2}}$ matrix.
\newblock {\em Physical Review B}, 102(4):041401, July 2020.
\newblock URL: \url{https://link.aps.org/doi/10.1103/PhysRevB.102.041401}, \href {https://doi.org/10.1103/PhysRevB.102.041401} {\path{doi:10.1103/PhysRevB.102.041401}}.

\bibitem{senadheera_hydrogen_2008}
Lasitha Senadheera and Mark~S. Conradi.
\newblock Hydrogen nuclear spin relaxation in hydrogen-ice clathrate.
\newblock {\em The Journal of Physical Chemistry A}, 112(36):8303--8309, September 2008.
\newblock URL: \url{https://pubs.acs.org/doi/10.1021/jp802858j}, \href {https://doi.org/10.1021/jp802858j} {\path{doi:10.1021/jp802858j}}.

\bibitem{mamone_theory_2011}
Salvatore Mamone, Judy Y.-C. Chen, Rangeet Bhattacharyya, Malcolm~H. Levitt, Ronald~G. Lawler, Anthony~J. Horsewill, Toomas Rõõm, Zlatko Bačić, and Nicholas~J. Turro.
\newblock Theory and spectroscopy of an incarcerated quantum rotor: {The} infrared spectroscopy, inelastic neutron scattering and nuclear magnetic resonance of {H}$_{\textrm{2}}$@{C}$_{\textrm{60}}$ at cryogenic temperature.
\newblock {\em Coordination Chemistry Reviews}, 255(7-8):938--948, April 2011.
\newblock URL: \url{https://linkinghub.elsevier.com/retrieve/pii/S0010854511000142}, \href {https://doi.org/10.1016/j.ccr.2010.12.029} {\path{doi:10.1016/j.ccr.2010.12.029}}.

\bibitem{mamone_rotor_2009}
S.~Mamone, Min Ge, D.~Hüvonen, U.~Nagel, A.~Danquigny, F.~Cuda, M.~C. Grossel, Y.~Murata, K.~Komatsu, M.~H. Levitt, T.~Rõõm, and M.~Carravetta.
\newblock Rotor in a cage: {Infrared} spectroscopy of an endohedral hydrogen-fullerene complex.
\newblock {\em The Journal of Chemical Physics}, 130(8):081103, February 2009.
\newblock URL: \url{https://pubs.aip.org/jcp/article/130/8/081103/922592/Rotor-in-a-cage-Infrared-spectroscopy-of-an}, \href {https://doi.org/10.1063/1.3080163} {\path{doi:10.1063/1.3080163}}.

\bibitem{yildirim_rotational_2002}
T.~Yildirim and A.~B. Harris.
\newblock Rotational and vibrational dynamics of interstitial molecular hydrogen.
\newblock {\em Physical Review B}, 66(21):214301, December 2002.
\newblock URL: \url{https://link.aps.org/doi/10.1103/PhysRevB.66.214301}, \href {https://doi.org/10.1103/PhysRevB.66.214301} {\path{doi:10.1103/PhysRevB.66.214301}}.

\bibitem{di_cataldo_giant_2024}
Simone Di~Cataldo, Maria Rescigno, Lorenzo Monacelli, Umbertoluca Ranieri, Richard Gaal, Stefan Klotz, Jacques Ollivier, Michael~Marek Koza, Cristiano De~Michele, and Livia~Eleonora Bove.
\newblock Giant splitting of the hydrogen rotational eigenenergies in the {C}$_{\textrm{2}}$ filled ice.
\newblock {\em Physical Review Letters}, 133(23):236101, December 2024.
\newblock URL: \url{https://link.aps.org/doi/10.1103/PhysRevLett.133.236101}, \href {https://doi.org/10.1103/PhysRevLett.133.236101} {\path{doi:10.1103/PhysRevLett.133.236101}}.

\bibitem{fitzgerald_quantum_2010}
S.~A. FitzGerald, J.~Hopkins, B.~Burkholder, M.~Friedman, and J.~L.~C. Rowsell.
\newblock Quantum dynamics of adsorbed normal- and para-{H}$_{\textrm{2}}$, {HD}, and {D}$_{\textrm{2}}$ in the microporous framework {MOF}-74 analyzed using infrared spectroscopy.
\newblock {\em Physical Review B}, 81(10):104305, March 2010.
\newblock URL: \url{https://link.aps.org/doi/10.1103/PhysRevB.81.104305}, \href {https://doi.org/10.1103/PhysRevB.81.104305} {\path{doi:10.1103/PhysRevB.81.104305}}.

\bibitem{fitzgerald_quantum_2008}
S.~A. FitzGerald, K.~Allen, P.~Landerman, J.~Hopkins, J.~Matters, R.~Myers, and J.~L.~C. Rowsell.
\newblock Quantum dynamics of adsorbed {H}$_{\textrm{2}}$ in the microporous framework {MOF}-5 analyzed using diffuse reflectance infrared spectroscopy.
\newblock {\em Physical Review B}, 77(22):224301, June 2008.
\newblock URL: \url{https://link.aps.org/doi/10.1103/PhysRevB.77.224301}, \href {https://doi.org/10.1103/PhysRevB.77.224301} {\path{doi:10.1103/PhysRevB.77.224301}}.

\bibitem{stoicheff_high_1957}
B.~P. Stoicheff.
\newblock High resolution raman spectroscopy of gases: {IX}. {Spectra} of {H2}, {HD}, and {D2}.
\newblock {\em Canadian Journal of Physics}, 35(6):730--741, June 1957.
\newblock URL: \url{https://cdnsciencepub.com/doi/10.1139/p57-079}, \href {https://doi.org/10.1139/p57-079} {\path{doi:10.1139/p57-079}}.

\bibitem{pauling_rotational_1930}
Linus Pauling.
\newblock The rotational motion of molecules in crystals.
\newblock {\em Physical Review}, 36(3):430--443, August 1930.
\newblock URL: \url{https://link.aps.org/doi/10.1103/PhysRev.36.430}, \href {https://doi.org/10.1103/physrev.36.430} {\path{doi:10.1103/physrev.36.430}}.

\bibitem{yildirim_quantum_2003}
Taner Yildirim and A.~B. Harris.
\newblock Quantum dynamics of a hydrogen molecule confined in a cylindrical potential.
\newblock {\em Physical Review B}, 67(24):245413, June 2003.
\newblock URL: \url{https://link.aps.org/doi/10.1103/PhysRevB.67.245413}, \href {https://doi.org/10.1103/PhysRevB.67.245413} {\path{doi:10.1103/PhysRevB.67.245413}}.

\bibitem{cross_does_2001}
R.~J. Cross.
\newblock Does {H}$_{\textrm{2}}$ rotate freely inside fullerenes?
\newblock {\em The Journal of Physical Chemistry A}, 105(28):6943--6944, July 2001.
\newblock URL: \url{https://pubs.acs.org/doi/10.1021/jp011054d}, \href {https://doi.org/10.1021/jp011054d} {\path{doi:10.1021/jp011054d}}.

\bibitem{matanovic_hydrogen_2012}
Ivana Matanović, Jonathan~L. Belof, Brian Space, Kaido Sillar, Joachim Sauer, Juergen Eckert, and Zlatko Bačić.
\newblock Hydrogen adsorbed in a metal organic framework-5: {Coupled} translation-rotation eigenstates from quantum five-dimensional calculations.
\newblock {\em The Journal of Chemical Physics}, 137(1):014701, July 2012.
\newblock URL: \url{https://pubs.aip.org/jcp/article/137/1/014701/191629/Hydrogen-adsorbed-in-a-metal-organic-framework-5}, \href {https://doi.org/10.1063/1.4730906} {\path{doi:10.1063/1.4730906}}.

\bibitem{ye_h2_2013}
Shufeng Ye, Minzhong Xu, Stephen FitzGerald, Kirill Tchernyshyov, and Zlatko Bačić.
\newblock H$_{\textrm{2}}$ in solid {C}$_{\textrm{60}}$: {Coupled} translation-rotation eigenstates in the octahedral interstitial site from quantum five-dimensional calculations.
\newblock {\em The Journal of Chemical Physics}, 138(24):244707, June 2013.
\newblock \_eprint: https://pubs.aip.org/aip/jcp/article-pdf/doi/10.1063/1.4811220/15463303/244707\_1\_online.pdf.
\newblock \href {https://doi.org/10.1063/1.4811220} {\path{doi:10.1063/1.4811220}}.

\bibitem{grazulis_crystallography_2012}
Saulius Gražulis, Adriana Daškevič, Andrius Merkys, Daniel Chateigner, Luca Lutterotti, Miguel Quirós, Nadezhda~R. Serebryanaya, Peter Moeck, Robert~T. Downs, and Armel Le~Bail.
\newblock Crystallography {Open} {Database} ({COD}): an open-access collection of crystal structures and platform for world-wide collaboration.
\newblock {\em Nucleic Acids Research}, 40(D1):D420--D427, January 2012.
\newblock Publisher: Oxford University Press (OUP).
\newblock URL: \url{https://academic.oup.com/nar/article-lookup/doi/10.1093/nar/gkr900}, \href {https://doi.org/10.1093/nar/gkr900} {\path{doi:10.1093/nar/gkr900}}.

\bibitem{kuhne_cp2k_2020}
Thomas~D. Kühne, Marcella Iannuzzi, Mauro Del~Ben, Vladimir~V. Rybkin, Patrick Seewald, Frederick Stein, Teodoro Laino, Rustam~Z. Khaliullin, Ole Schütt, Florian Schiffmann, Dorothea Golze, Jan Wilhelm, Sergey Chulkov, Mohammad~Hossein Bani-Hashemian, Valéry Weber, Urban Borštnik, Mathieu Taillefumier, Alice~Shoshana Jakobovits, Alfio Lazzaro, Hans Pabst, Tiziano Müller, Robert Schade, Manuel Guidon, Samuel Andermatt, Nico Holmberg, Gregory~K. Schenter, Anna Hehn, Augustin Bussy, Fabian Belleflamme, Gloria Tabacchi, Andreas Glöß, Michael Lass, Iain Bethune, Christopher~J. Mundy, Christian Plessl, Matt Watkins, Joost VandeVondele, Matthias Krack, and Jürg Hutter.
\newblock {CP2K}: {An} electronic structure and molecular dynamics software package - {Quickstep}: {Efficient} and accurate electronic structure calculations.
\newblock {\em The Journal of Chemical Physics}, 152(19), May 2020.
\newblock Publisher: AIP Publishing.
\newblock URL: \url{https://pubs.aip.org/jcp/article/152/19/194103/199081/CP2K-An-electronic-structure-and-molecular}, \href {https://doi.org/10.1063/5.0007045} {\path{doi:10.1063/5.0007045}}.

\bibitem{adamo_toward_1999}
Carlo Adamo and Vincenzo Barone.
\newblock Toward reliable density functional methods without adjustable parameters: {The} {PBE0} model.
\newblock {\em The Journal of Chemical Physics}, 110(13):6158--6170, April 1999.
\newblock URL: \url{https://pubs.aip.org/jcp/article/110/13/6158/476177/Toward-reliable-density-functional-methods-without}, \href {https://doi.org/10.1063/1.478522} {\path{doi:10.1063/1.478522}}.

\bibitem{kong_first-principles_2011}
Lingzhu Kong, Yves~J. Chabal, and David~C. Langreth.
\newblock First-principles approach to rotational-vibrational frequencies and infrared intensity for {H}$_{\textrm{2}}$ adsorbed in nanoporous materials.
\newblock {\em Physical Review B}, 83(12):121402, March 2011.
\newblock URL: \url{https://link.aps.org/doi/10.1103/PhysRevB.83.121402}, \href {https://doi.org/10.1103/PhysRevB.83.121402} {\path{doi:10.1103/PhysRevB.83.121402}}.

\bibitem{van_kranendonk_crystal_1962}
J.~Van~Kranendonk and H.P. Gush.
\newblock The crystal structure of solid hydrogen.
\newblock {\em Physics Letters}, 1(1):22--23, April 1962.
\newblock URL: \url{https://linkinghub.elsevier.com/retrieve/pii/0031916362902652}, \href {https://doi.org/10.1016/0031-9163(62)90265-2} {\path{doi:10.1016/0031-9163(62)90265-2}}.

\bibitem{pileio_relaxation_2010}
Giuseppe Pileio.
\newblock Relaxation theory of nuclear singlet states in two spin-1/2 systems.
\newblock {\em Progress in Nuclear Magnetic Resonance Spectroscopy}, 56(3):217--231, April 2010.
\newblock URL: \url{https://linkinghub.elsevier.com/retrieve/pii/S0079656509000958}, \href {https://doi.org/10.1016/j.pnmrs.2009.10.001} {\path{doi:10.1016/j.pnmrs.2009.10.001}}.

\bibitem{xu_coupled_2009}
Minzhong Xu, Francesco Sebastianelli, Brittney~R. Gibbons, Zlatko Bačić, Ronald Lawler, and Nicholas~J. Turro.
\newblock Coupled translation-rotation eigenstates of {H}$_{\textrm{2}}$ in {C}$_{\textrm{60}}$ and {C}$_{\textrm{70}}$ on the spectroscopically optimized interaction potential: {Effects} of cage anisotropy on the energy level structure and assignments.
\newblock {\em The Journal of Chemical Physics}, 130(22):224306, June 2009.
\newblock URL: \url{https://pubs.aip.org/jcp/article/130/22/224306/938040/Coupled-translation-rotation-eigenstates-of-H2-in}, \href {https://doi.org/10.1063/1.3152574} {\path{doi:10.1063/1.3152574}}.

\bibitem{kohama_rotational_2009}
Yoshimitsu Kohama, Takeshi Rachi, Ju~Jing, Zhaofei Li, Jun Tang, Ryotaro Kumashiro, Satoru Izumisawa, Hitoshi Kawaji, Tooru Atake, Hiroshi Sawa, Yasujiro Murata, Koichi Komatsu, and Katsumi Tanigaki.
\newblock Rotational sublevels of an ortho-hydrogen molecule encapsulated in an isotropic {C}$_{\textrm{60}}$ {Cage}.
\newblock {\em Physical Review Letters}, 103(7):073001, August 2009.
\newblock URL: \url{https://link.aps.org/doi/10.1103/PhysRevLett.103.073001}, \href {https://doi.org/10.1103/PhysRevLett.103.073001} {\path{doi:10.1103/PhysRevLett.103.073001}}.

\bibitem{fitzgerald_orientational_2015}
Stephen~A. FitzGerald, Christopher~T. Eckdahl, Cooper~S. McDonald, Jocienne~N. Nelson, Kai Shinbrough, Holden W.~H. Lai, and Jesse L.~C. Rowsell.
\newblock Orientational ortho-{H}$_{\textrm{2}}$ pair interactions in the microporous framework {MOF}-5.
\newblock {\em Physical Review B}, 92(13):134304, October 2015.
\newblock URL: \url{https://link.aps.org/doi/10.1103/PhysRevB.92.134304}, \href {https://doi.org/10.1103/PhysRevB.92.134304} {\path{doi:10.1103/PhysRevB.92.134304}}.

\bibitem{solodovnik_structure_2009}
A.~A. Solodovnik.
\newblock Structure of nitrous oxide-carbon dioxide alloys.
\newblock {\em Low Temperature Physics}, 35(4):339--342, April 2009.
\newblock URL: \url{https://pubs.aip.org/ltp/article/35/4/339/251452/Structure-of-nitrous-oxide-carbon-dioxide-alloys}, \href {https://doi.org/10.1063/1.3117965} {\path{doi:10.1063/1.3117965}}.

\bibitem{andrews_chemistry_1989}
Lester Andrews and Martin Moskovits.
\newblock {\em Chemistry and physics of matrix-isolated species}.
\newblock North-Holland, Amsterdam Oxford New York, 1989.

\bibitem{gush_infrared_1960}
H.~P. Gush, E.~J. Allin, H.~L. Welsh, and W.~F.~J. Hare.
\newblock The infrared fundamental band of liquid and solid hydrogen.
\newblock {\em Canadian Journal of Physics}, 38(2):176--193, February 1960.
\newblock URL: \url{https://cdnsciencepub.com/doi/10.1139/p60-020}, \href {https://doi.org/10.1139/p60-020} {\path{doi:10.1139/p60-020}}.

\bibitem{grinev_separation_2007}
Timur~A. Grinev, Alexei~A. Buchachenko, and Roman~V. Krems.
\newblock Separation of \textit{ortho}‐ and \textit{para}‐hydrogen in {Van} der {Waals} complex formation.
\newblock {\em ChemPhysChem}, 8(6):815--818, April 2007.
\newblock URL: \url{https://chemistry-europe.onlinelibrary.wiley.com/doi/10.1002/cphc.200700061}, \href {https://doi.org/10.1002/cphc.200700061} {\path{doi:10.1002/cphc.200700061}}.

\bibitem{buch_preferential_1993}
V.~Buch and J.~P. Devlin.
\newblock Preferential adsorption of ortho-{H}$_{\textrm{2}}$ with respect to para-{H}$_{\textrm{2}}$ on the amorphous ice surface.
\newblock {\em The Journal of Chemical Physics}, 98(5):4195--4206, March 1993.
\newblock URL: \url{https://pubs.aip.org/jcp/article/98/5/4195/933996/Preferential-adsorption-of-ortho-H2-with-respect}, \href {https://doi.org/10.1063/1.465026} {\path{doi:10.1063/1.465026}}.

\bibitem{hockey_weakly_2022}
Emily~K. Hockey, Korina Vlahos, Thomas Howard, Jessica Palko, and Leah~G. Dodson.
\newblock Weakly bound complex formation between {HCN} and {CH}$_{\textrm{3}}${Cl}: {A} matrix-isolation and computational study.
\newblock {\em The Journal of Physical Chemistry A}, 126(20):3110--3123, May 2022.
\newblock URL: \url{https://pubs.acs.org/doi/10.1021/acs.jpca.2c00716}, \href {https://doi.org/10.1021/acs.jpca.2c00716} {\path{doi:10.1021/acs.jpca.2c00716}}.

\bibitem{hockey_matrix-formation_2024}
Emily~K. Hockey, Nathan McLane, Korina Vlahos, Laura~M. McCaslin, and Leah~G. Dodson.
\newblock Matrix-formation dynamics dictate methyl nitrite conformer abundance.
\newblock {\em Journal of Chemical Physics}, 160(9), 2024.
\newblock URL: \url{https://pubs.aip.org/aip/jcp/article/160/9/094303/3268105/Matrix-formation-dynamics-dictate-methyl-nitrite}, \href {https://doi.org/10.1063/5.0188433} {\path{doi:10.1063/5.0188433}}.

\bibitem{escribano_crystallization_2013}
Rafael~M. Escribano, Guillermo~M. Muñoz~Caro, Gustavo~A. Cruz-Diaz, Yamilet Rodríguez-Lazcano, and Belén Maté.
\newblock Crystallization of {CO}$_{\textrm{2}}$ ice and the absence of amorphous {CO}$_{\textrm{2}}$ ice in space.
\newblock {\em Proceedings of the National Academy of Sciences}, 110(32):12899--12904, August 2013.
\newblock URL: \url{https://pnas.org/doi/full/10.1073/pnas.1222228110}, \href {https://doi.org/10.1073/pnas.1222228110} {\path{doi:10.1073/pnas.1222228110}}.

\bibitem{isokoski_highly_2013}
K.~Isokoski, C.~A. Poteet, and H.~Linnartz.
\newblock Highly resolved infrared spectra of pure {CO}$_{\textrm{2}}$ ice (15–75 {K}).
\newblock {\em Astronomy \& Astrophysics}, 555:A85, July 2013.
\newblock URL: \url{http://www.aanda.org/10.1051/0004-6361/201321517}, \href {https://doi.org/10.1051/0004-6361/201321517} {\path{doi:10.1051/0004-6361/201321517}}.

\bibitem{sebastianelli_hydrogen_2007}
Francesco Sebastianelli, Minzhong Xu, Yael~S. Elmatad, Jules~W. Moskowitz, and Zlatko Bačić.
\newblock Hydrogen molecules in the small dodecahedral cage of a clathrate hydrate: {Quantum} translation-rotation dynamics of the confined molecules.
\newblock {\em The Journal of Physical Chemistry C}, 111(6):2497--2504, February 2007.
\newblock URL: \url{https://pubs.acs.org/doi/10.1021/jp067318j}, \href {https://doi.org/10.1021/jp067318j} {\path{doi:10.1021/jp067318j}}.

\bibitem{grimme_consistent_2010}
Stefan Grimme, Jens Antony, Stephan Ehrlich, and Helge Krieg.
\newblock A consistent and accurate \textit{ab initio} parametrization of density functional dispersion correction ({DFT}-{D}) for the 94 elements {H}-{Pu}.
\newblock {\em The Journal of Chemical Physics}, 132(15), April 2010.
\newblock Publisher: AIP Publishing.
\newblock URL: \url{https://pubs.aip.org/jcp/article/132/15/154104/926936/A-consistent-and-accurate-ab-initio}, \href {https://doi.org/10.1063/1.3382344} {\path{doi:10.1063/1.3382344}}.

\end{thebibliography}

\end{document}


\preprint{APS/123-QED}

\title{\textit{Supplementary Information for} Environment-imposed selection rules for nuclear-spin conversion of H$_2$ in molecular crystals}

\author{Nathan McLane}
\email{nmclane@umd.edu}
\affiliation{%
 Institute for Physical Science and Technology, \\ University of Maryland, College Park, MD 20742, USA
}%

\author{LeAnh Duckett}
\affiliation{%
 Department of Chemistry and Biochemistry, \\ University of Maryland, College Park, MD 20742, USA
}%

\author{Leah G. Dodson}
\email{ldodson@umd.edu}
\affiliation{%
 Department of Chemistry and Biochemistry, \\ University of Maryland, College Park, MD 20742, USA
}%

\date{\today}

\maketitle

\section{Verification of \COO{} Crystallinity}

\begin{figure}[h]
\includegraphics[width=86mm]{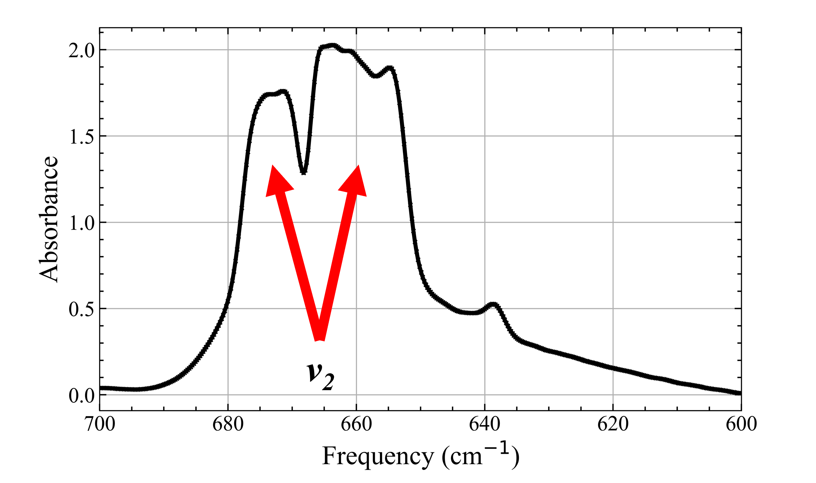}%
\caption{\label{fig:davy} \textbf{Verification of \COO{} Crystallinity} The crystallinity of the frozen molecular \COO{} ice is verified through the observation of Davydov splitting in the $\nu_2$ bending mode of \COO{} \cite{escribano_crystallization_2013,isokoski_highly_2013}.}
\end{figure}
\FloatBarrier

\newpage
\section{Potential energy plots and eigenfunctions for \HH{} in \NNO}

\begin{figure}[h]
\includegraphics[width=\linewidth]{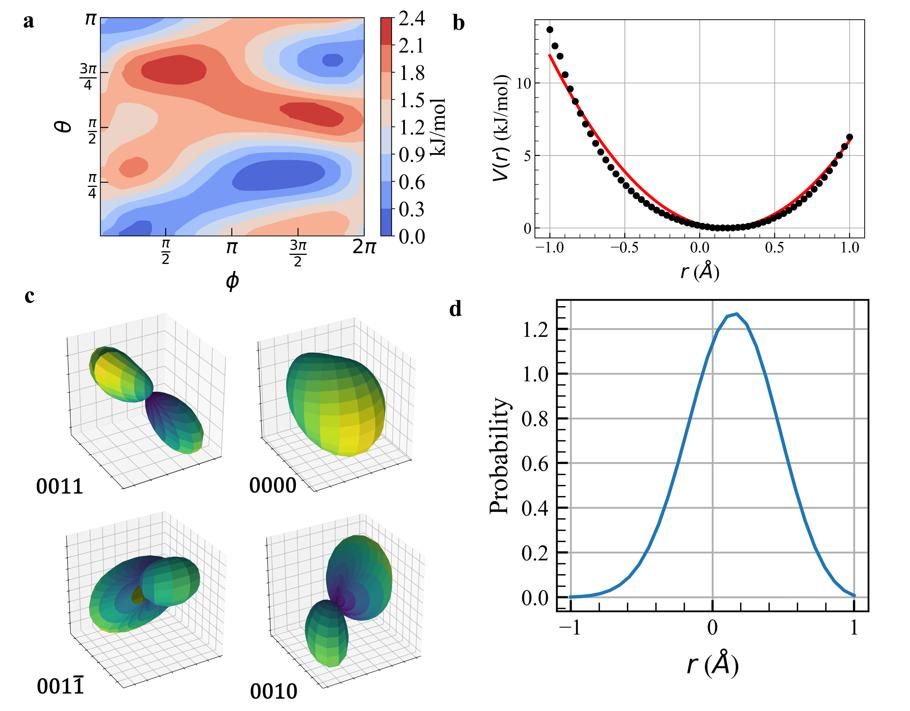}%
\caption{\label{fig:wfn-N2O} \textbf{Potential energy plots and eigenfunctions of \HH{} confined in solid \NNO{}.} \textbf{a,} Angular potential $V(\theta,\phi)$ for \HH{} in \NNO{} with respect to the two rotational degrees of freedom, $\theta$ and $\phi$. \textbf{b,} Radial potential $V(r)$ for \HH{} confined in \NNO{} as a function of the translational distance $r$ from equilibrium. The black dots are the calculated points and the red line is a fit assuming a harmonic potential. \textbf{c,} Reduced probability density plots of the angular wavefunctions of the four lowest-energy states (shown with state designations $nl\lambda m)$ resemble the spherical harmonics. \textbf{d,} The four lowest-energy translational eigenfunctions (shown) have identical probability densities stemming from the harmonic potential. }
\end{figure}

Experiments and theoretical analysis of \HH{} in \NNO{} provide verification for the tensor analysis of \HH{} trapped in non-magnetic hosts, Fig.~\ref{fig:wfn-N2O}. The same methods are followed as described in the main text for \COO{}, starting from the crystal structure of \NNO{} and substituting one \NNO{} unit with \HH{} \cite{grazulis_crystallography_2012,kuhne_cp2k_2020,adamo_toward_1999}.

\FloatBarrier

\newpage
\section{All computed eigenenergies}

Table 1 of the main text reports the eigenenergies obtained for the four lowest-energy states of \HH{} (one \pHH{} and three \oHH{}) trapped in \COO{} and \NNO{}. Given in Table \ref{tab:rotE} are the eigenenergies for the 20 lowest-energy states. 

\begin{table}[h]
\caption{\label{tab:rotE}%
Calculated energies (in \cm{}) of \HH{} in the perturbed system.
}
\begin{ruledtabular}
\begin{tabular}{ccc}
\multirow{2}{*}{State } & 
\multicolumn{2}{c}{Energy (cm$^{-1}$)} \\
\cline{2-3}
& \COO{} & \NNO{} \\
\hline
1 & 369.254444 & 284.2607947 \\
2 & 385.9940991 & 288.244126 \\
3 & 406.1376605 & 312.9363004 \\
4 & 430.0602659 & 322.2388535 \\
5 & 522.0499889 & 441.4159438 \\
6 & 539.6966095 & 445.399275 \\
7 & 558.9332054 & 470.0914494 \\
8 & 582.8558108 & 483.2687746 \\
9 & 602.9220299 & 512.0644716 \\
10 & 630.7166215 & 531.0030419 \\
11 & 659.5642193 & 543.928365 \\
12 & 675.7524994 & 546.7541936 \\
13 & 682.7128175 & 561.3182775 \\
14 & 687.6486137 & 602.4458648 \\
15 & 698.8665667 & 606.4291961 \\
16 & 712.6357158 & 631.1213705 \\
17 & 736.5583213 & 652.9628107 \\
18 & 755.7175749 & 669.2196206 \\
19 & 783.5121664 & 688.1581909 \\
20 & 812.3597642 & 701.0835141 \\
\end{tabular}
\end{ruledtabular}
\end{table}

\bibliographystyle{apsrev4-2}